\documentclass[twocolumn,prb,showpacs]{revtex4}
\usepackage{graphicx}
\usepackage{dcolumn}
\usepackage{bm}
\usepackage{amsmath}

\begin{document}

\preprint{}
\title{ Phonon Edelstein effect in chiral metals}
\author{Takehito Yokoyama}
\affiliation{Department of Physics, Institute of Science Tokyo, Tokyo 152-8551,
Japan
}
\date{\today}

\begin{abstract}
We propose a mechanism of current-induced phonon angular momentum, which we call phonon Edelstein effect. We investigate this effect in three-dimensional chiral metals with spin-orbit coupling and chiral phonons, and obtain an analytical expression of phonon angular momentum induced by the current. 
We also discuss the physical interpretation of this effect and give an estimation of its magnitude.
\end{abstract}

%\pacs{}
\maketitle

%\section{Introduction}

Phonons, the collective vibrational modes of atoms, are central to understanding the thermal, mechanical, and electronic properties of materials. In recent years, the discovery of chiral phonons --lattice vibrations with inherent handedness or rotational directionality-- has added a new dimension to the study of phonons.\cite{LZhang2014,LZhang2015,Zhang2023,Zhang2025}  Chiral phonons arise from the circular or elliptical motion of atoms in a crystal lattice, endowing these vibrational modes with angular momentum.\cite{HZhu2018,Ishito2023,Ishito2023b,Grissonnanche2020,XTChen2019,Li2019,Ueda2023,Zhang2024} This unique characteristic enables chiral phonons to interact with other quantum entities, such as electrons and spins, in ways that conventional phonons cannot.

The interplay between chiral phonons and spins has emerged as a promising area of research. Chiral phonons can directly influence spin dynamics through an effective magnetic field by chiral phonons.~\cite{Nova2017,Juraschek2019,Geilhufe,Juraschek2022,Xiong2022,Luo2023,Hernandez2022,Chaudhary2023} Recent researches in this field include conversion between electron spin and microscopic atomic rotation~\cite{HamadaPRR2020}, interaction between chiral phonons and magnons\cite{Yao2023,Wang2024,Ma2024,Fransson2025}, magnetization manipulation or reversal\cite{Basini,Kahana2023,Davies2023},
interactions between chiral phonons and eletronic spin\cite{Fransson2023} or orbital\cite{Ren2021} magnetizations, spin current generated by chiral phonons\cite{Kim2023,Li2022,Yao,Funato}, and  chiral phonon-mediated spin-spin interaction\cite{Korenev,Jeong2022,Yokoyama2024}. 
The coupling between chiral phonons and spins opens up exciting possibilities for designing novel spintronic devices, where phonon-mediated spin control could complement or even replace traditional methods reliant on external magnetic fields or spin-orbit coupling.

Phonon angular momentum can be induced by external means. 
Phonon angular momentum induced by the temperature gradient has been predicted.\cite{Hamada2018}  This is analogous to the Edelstein effect in electronic systems: current-induced spin angular momentum.\cite{Edelstein1990} 
Also, phonon angular momentum induced by electric field  in magnetic insulators has been predicted.\cite{Hamada2020}  This closely parallels magnetoelectric effects in insulators. 
In this paper, we further explore similarities with the Edelstein effect from a different perspective.

Also, spin polarization induced by charge current in chiral crystals have been experimentally investigated.\cite{Inui2020,Shiota2021} The spin signals are found over micrometer length scales which is much larger than spin diffusion length.\cite{Inui2020,Shiota2021}  The mechanism of this effect is still under debate. One possibility is that phonons carry angular momentum since phonon diffusion length is typically larger than spin diffusion length.
However, the mechism of current-induced phonon angular momentum still remains an open question.

In this paper,
we propose a mechanism of current-induced phonon angular momentum, which we call phonon Edelstein effect. We investigate this effect in three-dimensional chiral metals with spin-orbit coupling, in the presence of chiral phonons, and obtain an analytical expression of phonon angular momentum induced by the current. 
We also discuss the physical interpretation of this effect and give an estimation of its magnitude.

\begin{figure}[htb]
%\begin{figure*}[htb]
\begin{center}
\includegraphics[clip,width=5.0cm]{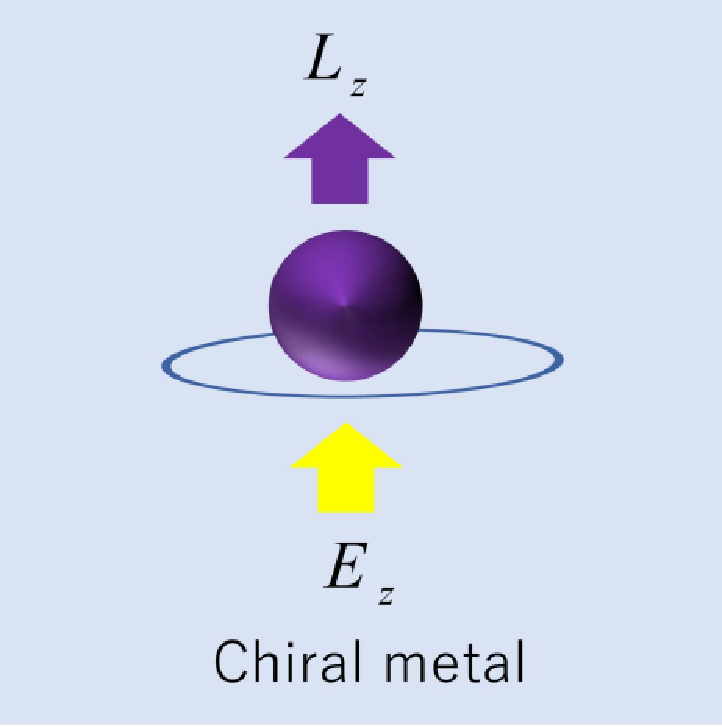}
\end{center}
\caption{
Schematic diagram of the model. The application of  an electric field to a chiral metal with chiral phonons can induce phonon angular momentum.}
\label{f1}
\end{figure}

We consider a three-dimensional chiral metal in the presence of chiral phonons and investigate  the phonon angular momentum in response to an electric field  as shown in Fig. \ref{f1}.
We assume that the chiral metal has a helical crystal structure with the space group $P3_1 21$ or $P3_2 21$ ($D^4_3$ or $D^6_3$) corresponding to the right-handed or left-handed screw symmetry. Te and Se have such a helical crystal structure.
 Then, the response tensor $\chi_{ij}$, defined by $L^i=\chi_{ij} E_{j}$, is diagonal: $\chi_{ij} \sim \delta_{ij}$\cite{Hamada2018,Birss1962} where $L^i$ and $E_{j}$ denote phonon angular momentum and an electric field, respectively. 
In this paper, specifically, we study the response of phonon angular momentum along the $z$-axis to an electric field applied along the $z$-axis.
In general, nonzero $\chi_{ij}$ is allowed for gyrotropic crystals. Among 21 point groups lacking inversion symmetry, 18 point groups are gyrotropic (from which 11 point groups are chiral).\cite{Kizel1975,Jerphagnon1976,Ganichev2019}
%---------------------------------------------------
%	Fig. 1: 
%---------------------------------------------------

We consider the Hamiltonian for phonons $H_p$ as follows
\begin{eqnarray}
{H_p} = \sum\limits_{{\bf{q}}\nu} {\omega _{\bf{q}\nu}} a_{\bf{q}}^{\nu\dag }a_{\bf{q}}^{\nu }
\end{eqnarray}
where $a_{\bf{q}}^{\nu \dag } (a_{\bf{q}}^{\nu})$ is the creation (annihilation) operator for a phonon with wave vector $\bf{q}$ and mode $\nu$, with the corresponding phonon dispersion $\omega _{\bf{q}\nu}$.
We set $k_B=\hbar=1$ throughout this paper. 

We define the phonon Green's function $D$ by
\begin{eqnarray}
D({\bf{q}},t ,\nu) =  - i\left\langle {{{\rm{T}}_\tau }\left[ {{\varphi _{{\bf{q}}\nu}}(t)\varphi _{{\bf{q}}\nu}^\dag } \right]} \right\rangle 
\end{eqnarray}
with $\varphi _{{\bf{q}}\nu}^\dag  = \frac{1}{{ 2 }}\left( {a_{\bf{q}}^\nu + a_{ - {\bf{q}}}^{\nu\dag }, - i\left( {a_{\bf{q}}^{\nu } - a_{ - {\bf{q}}}^{\nu \dag} } \right)} \right)$.
These two components correspond to the second quantization representation of the displacement and momentum of atoms.

We perform a Fourier transform of the phonon Green's function $D({\bf{q}},t ,\nu)$ for $H_p$ and obtain the retarded phonon Green's function:
\begin{widetext}
\begin{eqnarray}
{D^r}({\bf{q}},\omega ,\nu) = \frac{1}{{2\left[ {(\omega  + i/{\tau _p})_{}^2 - \omega _{{\bf{q}}\nu}^2} \right]}}\left( {\begin{array}{*{20}{c}}
{\omega _{{\bf{q}}\nu}^{}}&{i(\omega  + i/{\tau _p})}\\
{ - i(\omega  + i/{\tau _p})}&{\omega _{{\bf{q}}\nu}^{}}
\end{array}} \right) \nonumber \\
\equiv D_0^r({\bf{q}},\omega ,\nu){\tau _0} + D_2^r({\bf{q}},\omega ,\nu){\tau _y}
\end{eqnarray}
\end{widetext}
with the phonon relaxation time $\tau _p$ and the Pauli matrices $\tau_0$ and $\tau_y$.
The angular momentum of phonons along the $z$ direction can be expressed by the second quantization representation of the displacement $u$ and momentum $p$  as \cite{Gao2023}
\begin{eqnarray}
{L^z} = u^xp^y - u^yp^x = \frac{1}{{2}}\sum\limits_{{\bf{q}},\nu} {g_{\bf{q}}^{\nu ,{\nu^\prime }}\left( {a_{\bf{q}}^\nu + a_{ - {\bf{q}}}^{\nu\dag }} \right)} \left( {a_{\bf{q}}^{\nu '\dag } - a_{ - {\bf{q}}}^{\nu'}} \right)
\end{eqnarray}
with
\begin{eqnarray}
g_{\bf{q}}^{\nu ,{\nu^\prime }} = \sqrt {\frac{{\omega _{\bf{q}\nu^\prime }}}{{\omega _{\bf{q}\nu}}}} \xi _{{\bf{q}},{\nu ^\prime }}^\dag \left( {\begin{array}{*{20}{c}}
0&{ - i}\\
i&0
\end{array}} \right){\xi _{{\bf{q}},\nu }}.
\end{eqnarray}
Here, ${\xi _{{\bf{q}},\nu }}$ is the polarization vector. We assume two oppositely circularly polarized phonon modes which may be realized in hexagonal lattices.\cite{LZhang2015} 
%These modes are in general not circularly polarized. However, we can obtain circular polarized phonon modes by superposition of the degenerated modes.\cite{LZhang2014}  
Each of circular polarized phonon modes is labeled by $\nu=\pm$. Namely, we assume
${\xi _{{\bf{q}},\nu }} = \frac{1}{{\sqrt 2 }}{\left( {\begin{array}{*{20}{c}}
1&{\nu  i} \end{array}} \right)^t}$.
Then we have
$g_{\bf{q}}^{\nu ,{\nu ^\prime }} = \nu {\delta _{\nu ,\nu '}}$
and obtain
\begin{eqnarray}
{L^z} = \sum\limits_{{\bf{q}},\nu } {\nu \varphi _{{\bf{q}}\nu }^\dag {\tau_y}\varphi _{{\bf{q}}\nu }^{}}.
\end{eqnarray}

We now turn to the description of the electronic system.
We consider the model Hamiltonian for electrons in a chiral metal in matrix form  $H_e$ as
\begin{eqnarray}
H_e = \frac{{{k^2}}}{{2m}} + \alpha {\bf{k}} \cdot {\bm{\sigma }}
\end{eqnarray}
where ${\bf{k}}$ and $\bm{\sigma }$ denote wavevector and spin Pauli matrix of electrons, respectively. The second term in this Hamiltonian represents Weyl type spin-orbit coupling and breaks inversion and mirror symmetries.\cite{Burkov2018,Armitage2018}
This type of Hamiltonian can describe low energy physics of chiral crystals such as tellurium.\cite{Calavalle2022}

We consider the  interaction between  chiral phonons and electrons in the form of spin-orbit coupling:
\begin{eqnarray}
{H_{ep}} = \lambda {L^z}{\sigma _z}.
\end{eqnarray}
The parameter $\lambda$ represents the coupling strength.
This interaction stems from Zeeman effect by the magnetic field due to chiral phonons. We assume that  atoms rotate within the $xy$ plane. Then, the associated magnetic field is parallel to the $z$ axis and hence only $\sigma _z$ is involved in this couplig.

The applied electric field is also described by 
\begin{eqnarray}
{H_A} =  - {j_z}{A_z}
\end{eqnarray}
where the current operator $j_z$ and the vector potential $A_z$ are, respectively, given by 
\begin{eqnarray}
{j_z} =  - e\left( {\frac{{{k_z}}}{m} + \alpha {\sigma _z}} \right), \; {A_z} =  - \frac{i}{\Omega }{E_z} =  - \frac{i}{\Omega }{E_{0z}}{e^{ - i\Omega t}}.
\end{eqnarray}
Here, $E_z$ is the electric field with the frequency $\Omega $ and a constant $E_{0z}$.

The electronic retarded Green's functions for $H_e$ are obtained as 
\begin{eqnarray}
G_{{\bf{k}}\omega }^r = (\omega  + i/{\tau _e} - {H_e})\nonumber \\  \approx \frac{1}{{\omega  + i/{\tau _e} - {\varepsilon _k}}} - \frac{{\alpha {\bf{k}} \cdot {\bm{\sigma }}}}{{{{\left( {\omega  + i/{\tau _e} - {\varepsilon _k}} \right)}^2}}}\nonumber \\ \equiv G_{0{\bf{k}}\omega }^r + {\bf{G}}_{{\bf{k}}\omega }^r \cdot {\bm{\sigma }}.
\end{eqnarray}
with $\varepsilon _k=\frac{{{k^2}}}{{2m}}$. Here, we expand the Green's functions up to first order of $\alpha$.
 $\tau _e$ is the electronic relaxation time.

In the following, we treat ${H_{ep}} $ and  ${H_A} $ as a perturbation.
We calculate the expectation value of ${L^z}$ with respect to the first orders of ${H_{ep}} $ and  ${H_A} $, which can be expressed by the lesser phonon Green's function:
\begin{eqnarray}
\left\langle {{L^z}} \right\rangle  =   i{\rm{Tr}}{\nu \tau _y}{\cal{D}^ < }\nonumber \\  \approx -i{\rm{Tr}} {\left[ {\nu{\tau _y}{D_{{\bf{q}}\omega \nu}}\lambda \nu{\tau _y}{\sigma _z}{G_{{\bf{k}}\omega '}}{j_z}{A_z}{G_{{\bf{k}}\omega ' + \Omega }}{D_{{\bf{q}}\omega \nu}}} \right]^ < }.
\end{eqnarray}
Here, $\cal{D}$ is the full phonon Green's function. The trace is taken over the phononic and electronic degrees of freedom. The diagrammatic representations of this expansion of the phonon Green's function  are shown in Fig. 2.

%---------------------------------------------------
%	Fig. 2: 
%---------------------------------------------------
\begin{figure}[htb]
%\begin{figure*}[htb]
\begin{center}
\includegraphics[clip,width=8cm]{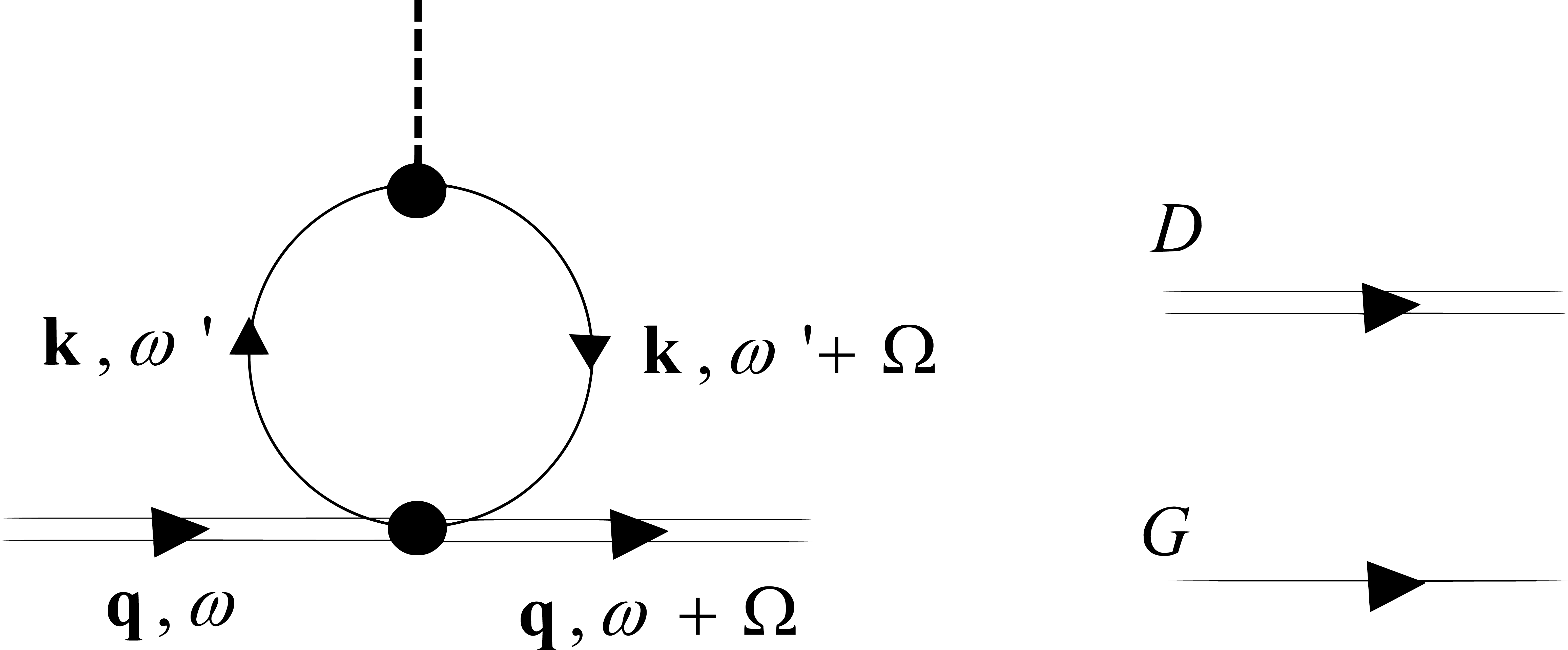}
\end{center}
\caption{
The diagrammatic representations of the phonon Green's function to first orders in ${H_{ep}} $ and  ${H_A} $. Two parallel solid, solid and dotted lines represent phonon Green's function, electron Green's function, and ${H_A} $, respectively.  Two parallel solid and solid lines intersect via ${H_{ep}} $.  }
\label{f2}
\end{figure}
%\end{figure*}
%---------------------------------------------------

We expand the lesser component by the Langreth theorem\cite{HaugJauho1997}.
Since we consider the current-induced phonon angular momentum under the metallic condition $\varepsilon _F \tau_e \gg 1$, we focus on the lesser Green's function for electrons:
\begin{widetext}
\begin{eqnarray}
\left\langle {{L^z}} \right\rangle  \approx -i\lambda {\rm{Tr}}\left[ {{\tau _y}D_{{\bf{q}}\omega v}^r{\tau _y}{\sigma _z}\left( {G_{{\bf{k}}\omega '}^r{j_z}{A_z}G_{{\bf{k}}\omega ' + \Omega }^ <  + G_{{\bf{k}}\omega '}^ < {j_z}{A_z}G_{{\bf{k}}\omega ' + \Omega }^a} \right)D_{{\bf{q}}\omega v}^a} \right].
\end{eqnarray}
Noting that
$
G^{<}_{\mathbf{k}\omega} = f_{\omega}\left( G^{a}_{\mathbf{k}\omega} -  G^{r}_{\mathbf{k}\omega} \right),
$
where $f_{\omega}$ is the Fermi distribution function, we have at zero temperature
\begin{eqnarray}
G_{{\bf{k}}\omega '}^r{j_z}{A_z}G_{{\bf{k}}\omega ' + \Omega }^ <  + G_{{\bf{k}}\omega '}^ < {j_z}{A_z}G_{{\bf{k}}\omega ' + \Omega }^a \approx \Omega f'_{\omega'}G_{{\bf{k}}\omega '}^r{j_z}{A_z}G_{{\bf{k}}\omega '}^a \nonumber \\
 =  - \Omega \delta (\omega ' - {\varepsilon _F})G_{{\bf{k}}\omega '}^r{j_z}{A_z}G_{{\bf{k}}\omega '}^a
\end{eqnarray}
where $\varepsilon _F$ is the Fermi energy.

We now take trace over the electronic degrees of freedom:
\begin{eqnarray}
{\rm{T}}{{\rm{r}}_{\sigma {\bf{k}}\omega '}}{\sigma _z}\left( {G_{{\bf{k}}\omega '}^r{j_z}{A_z}G_{{\bf{k}}\omega ' + \Omega }^ <  + G_{{\bf{k}}\omega '}^ < {j_z}{A_z}G_{{\bf{k}}\omega ' + \Omega }^a} \right) =  - \frac{{ie{E_z}}}{\pi }\sum\limits_{\bf{k}} {\left( {G_z^r\frac{{{k_z}}}{m}G_0^a + G_0^a\frac{{{k_z}}}{m}G_z^r + \alpha G_0^rG_0^a} \right)}.
\end{eqnarray}
Here and in the following, we omit the subscripts ${\bf{k}}$ and $\omega$ of the Green's functions for notational simplicity. The subscripts of Tr indicate the degrees of freedom over which the partial trace is taken.
Each term is calculated as 
\begin{eqnarray}
\sum\limits_{\bf{k}} {\left( {G_z^r\frac{{{k_z}}}{m}G_0^a + G_0^a\frac{{{k_z}}}{m}G_z^r} \right)}  =  - \frac{{2\alpha {\nu _e}}}{3}\int_{}^{} {d{\varepsilon _k}\frac{1}{{{{\left( {{\varepsilon _F} + i/{\tau _e} - {\varepsilon _k}} \right)}^2}\left( {{\varepsilon _F} - i/{\tau _e} - {\varepsilon _k}} \right)}}}  + c.c. = \frac{{2\pi \alpha {\nu _e}{\tau _e}}}{3},
\end{eqnarray}

\begin{eqnarray}
\sum\limits_{\bf{k}} {\left( {G_0^rG_0^a} \right)}  = {\nu _e}\int_{}^{} {d{\varepsilon _k}\frac{1}{{{{\left( {{\varepsilon _F} - {\varepsilon _k}} \right)}^2} + {{(1/{\tau _e})}^2}}}}  = \pi {\nu _e}{\tau _e}.
\end{eqnarray}
Here, $\nu _e$ is the density of states of electrons at the Fermi energy. 

As for the phonon part, since ${\rm{T}}{{\rm{r}}_{\tau {\bf{q}}\omega }}{\tau _y}D_{{\bf{q}}\omega \nu}^r{\tau _y}D_{{\bf{q}}\omega \nu}^a = 2{\rm{T}}{{\rm{r}}_{{\bf{q}}\omega }}D_0^rD_0^a + D_2^rD_2^a$, we have
\begin{eqnarray}
\sum\limits_\omega  {\left( {D_0^rD_0^a + D_2^rD_2^a} \right)}  = \frac{1}{{2\pi }}\int_{}^{} {d\omega \frac{{\omega _{{\bf{q}}\nu}^2 + {\omega ^2} + {{\left( {1/{\tau _p}} \right)}^2}}}{{\left[ {{{\left( {\omega  + i/{\tau _p}} \right)}^2} - \omega _{{\bf{q}}\nu}^2} \right]\left[ {{{\left( {\omega  - i/{\tau _p}} \right)}^2} - \omega _{{\bf{q}}\nu}^2} \right]}}}  = \frac{{{\tau _p}}}{8}.
\end{eqnarray}
\end{widetext}

Therefore, we obtain the final expression: 
\begin{eqnarray}
\left\langle {{L^z}} \right\rangle  = \chi {E_z},\; \chi  = -\frac{5}{{12}}e\alpha \lambda {\nu _e}{\tau _e}{\tau _p}
\end{eqnarray}
which exhibits the current-induced phonon angular momentum.

The physical interpretation of this effect is as follows.
According to the Edelstein effect, the spin polarization is induced by an applied electric field, i.e., $\left\langle {{\sigma _z}} \right\rangle \sim E$.\cite{Edelstein1990}
By the mean field approximation, the electron-chiral phonon coupling becomes 
\begin{eqnarray}
{H_{ep}} = \lambda {L^z}\left\langle {{\sigma _z}} \right\rangle.
\end{eqnarray}
Therefore, the phonon angular momentum stems from an orbital Zeemann effect with an effective magnetic field $\left\langle {{\sigma _z}} \right\rangle$ which is proportional to the electric field.

We now estimate the magnitude of $\left\langle {{L^z}} \right\rangle $ in Eq.(19). For $\tau_p \sim 10^{-10}$s, $\tau_e \sim 10^{-12}$s, $E_z \sim 10^{4}$V/m, $\lambda \sim 1$meV,\cite{Gao2023} $\nu_e \sim 1$/(eV$\cdot$unit cell), and
$\alpha \sim 1$eV $\cdot$ \AA, we have $\left\langle {{L^z}} \right\rangle  \sim 10^{-1} \hbar$/unit cell.
For the conductivity $10^3$ S/m, the corresponding current density is $10^7$ A/m$^2$.
For comparison, the magnitude of phonon angular momentum induced by the temperature gradient is estimated as $10^{-7} \hbar$/\AA$^3$ at room temperature under the temperature gradient 10K/$\mu$m for  $\tau_p \sim 10^{-11}$s.\cite{Hamada2018}
The magnitude of phonon angular momentum induced by electric field in magnetic
insulators is estimated as $10^{-8} \hbar$ per unit cell at room temperature.\cite{Hamada2020}
As for experimental signatures of this effect, we propose that by examining the difference in circularly polarized Raman scattering with and without an applied current, one can detect the current-induced phonon angular momentum.

Let us discuss a contribution from electronic orbital angular momentum. 
The orbital magnetic moment of a Bloch electron can be calculated as \cite{Chang1996,Sundaram1999,Xiao2010}
\begin{equation}
{\bm{L}}_{}^e=-m \operatorname{Im}\left\langle\partial_{\mathbf{k}} u_{\mathbf{k} n}\right| \times\left(H_e-\epsilon_{\mathbf{k} n}\right)\left|\partial_{\mathbf{k}} u_{\mathbf{k} n}\right\rangle .
\end{equation}
Here, $\left|u_{\mathbf{k} n}\right\rangle$ and $\epsilon_{\mathbf{k} n}$ denote the Bloch states and the band energy, respectively. $n$ labels the bands.
For Weyl type Hamiltonians as Eq.(7), this orbital magnetic moment has been calculated in \cite{Zhong2016}.
The orbital magnetic moment induced by the electric  field along the $z$ axis is obtained as \cite{Zhong2016}
\begin{equation}
\left\langle {L_z^e} \right\rangle  =  - \frac{2}{3}em\alpha {k_F}{\tau _e}{E_z}.
\end{equation}
Here, $k_F$ denotes the Fermi wavenumber. Now, we consider the interaction between chiral phonons and electrons of the form:
\begin{eqnarray}
{H_{ep}'} = \lambda {L^z}{L_z^e}.
\end{eqnarray}
This interaction represents the orbital Zeeman coupling between the magnetic field due to chiral phonons and electronic orbital angular momentum, which leads to the generation of phonon angular momentum.
By considering ${H_{ep}^{'}}$ instead of  ${H_{ep}}$, we can calculate the expectation value of the phonon angular momentum $L^{z'}$ in a similar way. 
We obtain
\begin{equation}
\left\langle {L^{z'}} \right\rangle  = {\chi _o}{E_z}, \; {\chi _o} = \frac{{m{k_F}}}{{5{\nu _e}}}\chi.
\end{equation}
For $\frac{{k_F^2}}{{2m}} \sim$ 1eV and $\nu_e \sim 1$/(eV$\cdot$unit cell), we have ${\chi _o}/\chi  \sim 0.1$.
Thus, the contribution from the electronic orbital angular momentum is comparable or smaller than that from  the electronic spin angular momentum.

Recently, chiral phonons have been observed in a chiral crystal tellurium.\cite{Ishito2023b}
Since tellurium is a Weyl semiconductor with strong spin-orbit coupling\cite{Zhang2020,Qiu2020}, doped tellurium is a promising candidate material to verify our prediction.

Finally, let us comment on the reciprocal effect: inverse phonon Edelstein effect or magnetic field-induced charge current.
If we apply an magnetic field along the $z$-axis $B_z$ and consider an orbital Zeemann coupling $H_B=L^z B_z$ instead of $H_A$, we obtain charge current along the $z$-axis in response to the magnetic field according to the Onsager reciprocity. 
The current induced by a slowly-varying magnetic field has been formulated in terms of the intrinsic magnetic moment of the Bloch states on the Fermi surface, which is dubbed gyrotropic magnetic effect.\cite{Zhong2016}
The inverse phonon Edelstein effect proposed here can be interpreted as chiral phonon-induced gyrotropic magnetic effect.

In summary, 
we have proposed a mechanism of current-induced phonon angular momentum, i.e., phonon Edelstein effect. We have investigated this effect in three-dimensional chiral metals in the presence of spin-orbit coupling and chiral phonons, and obtained an analytical expression of phonon angular momentum induced by the current.  We have also discussed the physical interpretation of this effect and given an estimation of its magnitude.

This work was supported by JSPS KAKENHI Grant No.~JP19K03712 and JP25K07221.

\end{document}